\documentclass[12pt]{iopart}

\usepackage{graphicx}
\begin{document}

\title{Is there a quad problem among optical gravitational lenses?}

\author{Masamune Oguri}
\address{Kavli Institute for Particle Astrophysics and
Cosmology, Stanford University, 2575 Sand Hill Road, Menlo Park, 
CA 94025, USA}
\ead{oguri@slac.stanford.edu}
\begin{abstract}
Most of optical gravitational lenses recently discovered in the Sloan
Digital Sky Survey Quasar Lens Search (SQLS) have two-images rather
than four-images, in marked contrast to radio lenses for which the
fraction of four-image lenses (quad fraction) is quite high. We
revisit the quad fraction among optical lenses by taking the selection
function of the SQLS into account. We find that the current observed
quad fraction in the SQLS is indeed lower than, but consistent with, the
prediction of our theoretical model. The low quad fraction among
optical lenses, together with the high quad fraction among radio
lenses, implies that the quasar optical luminosity function has a
relatively shallow faint end slope.   
\end{abstract}

\maketitle

\section{Introduction}

Strongly lensed multiple quasars have been known to provide an unique
probe of our universe. In particular, the point-source nature of quasars
allows a simple statistical study from image multiplicities: Statistics
of the number of multiple images provide constraints on the ellipticity
and density profile of lens objects as well as the faint end luminosity
function of source quasars 
\cite{blandford87,kormann94,kochanek96,keeton97,rusin01,chae03,cohn04,oguri04a,huterer05}.

The statistics of image multiplicities have been done mainly using radio
lenses. \cite{rusin01} adopted a radio lens sample of the Cosmic Lens
All-Sky Survey (CLASS) \cite{myers03,browne03} to show that the fraction
of four-image (quadruple) lenses is significantly
higher than expected from a standard mass model of elliptical
galaxies. \cite{chae03} showed that the fraction of quadruple lenses
in a statistical subsample of the CLASS is marginally consistent with
what we expect from the observed galaxy population, but it still
requires relatively large galaxy ellipticities.  

Recent large-scale optical surveys allow us to conduct complementary
statistics using optical gravitational lenses. In particular, a large sample
of quasars discovered in the Sloan Digital Sky Survey (SDSS)
\cite{york00} is quite useful for a strong lens survey: Indeed, the SDSS
Quasar Lens Search (SQLS) \cite{oguri06b} has already discovered
approximately 20 new strongly lensed quasars (see, e.g., \cite{inada07}
and references therein), becoming the largest statistical sample of
strongly lensed quasars. Interestingly, the fraction of four-image lenses
(quad fraction) in the SQLS appears to be significantly lower than the
CLASS. Only a few lenses among $\sim20$ new SQLS lensed quasars are
quadruple lenses, whereas nearly half of CLASS lenses were four (or
more) image systems.  

In this paper, we revisit the quad fraction among optical gravitational
lenses. We adjust the selection function to that of the SQLS and make a
comprehensive prediction of the fraction of quadruple lenses. A
particular emphasis is paid to whether the current low quad fraction
in the SQLS is consistent with the observed galaxy properties.  
Throughout the paper we adopt $\Lambda$-dominated cosmology with the
matter density $\Omega_M=0.3$ and the cosmological constant
$\Omega_\Lambda=0.7$. 

\section{Calculation}

\subsection{Lensing Probabilities}
We assume that the mass distribution of galaxies can be approximated by
an Singular Isothermal Ellipsiod (SIE). The scaled surface mass density
of an SIE is given by
\begin{equation}
\kappa(x,y)=\frac{\theta_{\rm E}\lambda(e)}{2}\left[\frac{1-e}{(1-e)^2x^2+y^2}\right]^{1/2},
\end{equation}
where $e$ denotes the ellipticity. The Einstein radius $\theta_{\rm E}$
(for $e=0$) is related with the galaxy velocity dispersion $\sigma$ by
\begin{equation}
\theta_{\rm E}=4\pi\left(\frac{\sigma}{c}\right)^2\frac{D_{\rm ls}}{D_{\rm os}},
\end{equation}
with $D_{\rm ls}$ and $D_{\rm os}$ being the angular diameter distance
from lens to source and from observer to source, respectively. 
The normalization factor $\lambda(e)$ basically depends on the shape
and viewing angle of galaxies: In this paper we assume that
there are equal number of oblate and prolate galaxies and adopt the
average of the two normalizations (see \cite{chae03}). We find that with
this normalization the Einstein radii are roughly equal for different
ellipticities.  

It is expected that the quad fraction is mainly determined by the
ellipticity. Although the external shear also produces the quadrupole
moment in lens potentials, the effect is expected to be minor. For
instance, the standard strength of external shear (median value of
$<0.05$)  can cause notable changes in the quad fraction only for lens
galaxies with $e<0.2$ \cite{rusin01}. Therefore throughout the paper
we neglect the external shear. 

We solve the lens equation using a public code {\it lensmodel}
\cite{keeton01}. The lensing cross section $\sigma_{\rm lens}$ is
computed by summing up source positions that yield multiple images with
a weight of $\Phi(L/\mu)/\mu/\Phi(L)$, where $\Phi(L)$ is the luminosity
function of source quasars and $\mu$ is the magnification factor (see \S
\ref{sec:qso} for which magnification factor we adopt). Lensing cross
sections are derived for double and quadruple lenses separately. We 
compute the image separation for each event from the maximum separation
between any image pairs. In computing the lensing probability, we
impose a condition that the lensing 
galaxy should not be brighter than the source quasar, because the
lens system may not be targeted for spectroscopy if the lensing
galaxy dominates in the flux. We compute the galaxy luminosity from
the velocity dispersion adopting an observed correlation
\cite{bernardi03}.  Then the lensing probability of a source at
$z=z_s$ becomes
\begin{equation}
\frac{dp_i}{d\theta}=\int_0^{z_s}dz_l 
 \frac{c\,dt}{dz_l}(1+z_l)^3\int d\sigma \frac{d\sigma_{{\rm
 lens},i}}{d\tilde{\theta}}\frac{dn}{d\sigma}\delta(\theta-\tilde{\theta})
 \Theta(i_{\rm gal}-i_{\rm qso}),
\end{equation}
with $dn/d\sigma$ being the velocity function of galaxies. The suffix
$i=2$ or $4$ denote the number of images. 

In computing the lensing probability, we need to specify the lens galaxy
population. Since strong lensing is  mostly caused by early-type
galaxies, particularly for strong lenses in the SQLS whose image
separations are basically larger than $1''$, we only consider
early-type galaxies. For the velocity function, we assume that of
early-type galaxies derived from the SDSS \cite{choi07,chae07}. More
important for the quad fraction is the distribution of ellipticities. 
We adopt a Gaussian distribution with mean $\bar{e}=0.3$ and the
dispersion $\sigma_e=0.16$, which is consistent with observed
ellipticity distributions of early-type galaxies
\cite{bender89,saglia93,jorgensen95,rest01,sheth03}, as a fiducial
distribution. However we also vary the mean ellipticity, $\bar{e}$, to
see how the quad fraction depends on the ellipticity.

\subsection{Quasar Population and Selection Function}\label{sec:qso}
The quasar luminosity function is another important element to make an
accurate prediction of the quad fraction. As a fiducial luminosity
function, we adopt that constrained from the combination of the SDSS
and 2dF \cite{richards05}:
\begin{equation}
\Phi(M_g)=\frac{\Phi_*}{10^{0.4(1-\beta_{\rm h})(M_g-M_g^*)}+10^{0.4(1-\beta_{\rm l})(M_g-M_g^*)}},
\end{equation}
where a pure luminosity evolution with 
\begin{equation}
M_g^*(z)=M_g^*(0)-2.5(k_1z+k_2z^2)
\end{equation}
is assumed. The parameters are $\beta_{\rm h}=3.31$, $\beta_{\rm
  l}=1.45$, $\Phi_*=1.83\times10^{-6}{\rm Mpc}^{-3}{\rm mag}^{-1}$, 
$M_g^*(0)=-21.61$, $k_1=1.39$, and $k_2=-0.29$. We convert 
 rest-frame $g$-band magnitudes to observed $i$-band magnitudes using
 K-correction derived in \cite{richards06}.

The selection function of the SQLS was studied in detail in
\cite{oguri06b}. Since the statistical sample of lensed quasars is
constructed from quasars with $i<19.1$ and at $0.6<z<2.2$, we restrict
our calculation in this range. The magnification bias is computer
assuming the image separation dependent magnification factor (see
equation (14) in \cite{oguri06b}). At $\theta>1''$ the completeness is
almost unity, but there is a small difference of completeness between
double and quad lenses: To take this into account we include
completeness $\phi_i(\theta)$ in our calculation. In summary, we
compute the numbers of double and quad lenses as
\begin{equation}
n_i=\int_{0.6}^{2.2} dz_s\int_{i<i_{\rm lim}} dM_g \Phi(M_g) 
\Omega D_{\rm os}^2
\frac{c\,dt}{dz_s}(1+z_s)^3 \int_{1''}^{3''}d\theta \phi_i(\theta)
\frac{dp_i}{d\theta},
\end{equation}
where $i_{\rm lim}=19.1$ for the statistical lens sample of the
SQLS. We have set the upper limit of the image separation to $3''$
since beyond the image separation the effect of surrounding dark
matter becomes significant (see, e.g., \cite{oguri06a}). The fraction
of quadruple lenses is then computed as  
\begin{equation}
p_Q=\frac{n_4}{n_2+n_4}.
\end{equation}

\subsection{Lensed Quasars in the SQLS}

\begin{table}
\caption{A current statistical sample of lensed quasars in the 
 SQLS. $N_{\rm img}$ indicate the number of quasar images. \label{tab:sdss}}  
\begin{indented}
\lineup
\item[]\begin{tabular}{@{}*{4}{l}}
\br                              
Name&$N_{\rm img}$&$i_{\rm PSF}$&Ref.\cr 
\mr
SDSS J0246$-$0825 &2 &17.8 &\cite{inada05}\cr 
SDSS J0746+4403 &2 &18.8   &\cite{inada07}\cr 
SDSS J0806+2006 &2 &19.0   &\cite{inada06}\cr 
SBS0909+523     &2 &16.2   &\cite{oscoz97}\cr 
SDSS J0924+0219 &4 &18.2   &\cite{inada03a}\cr 
FBQ0951+2635    &2 &17.3   &\cite{schechter98}\cr 
SDSS J1001+5027 &2 &17.3   &\cite{oguri05}\cr 
SDSS J1021+4913 &2 &19.0   &\cite{pindor06}\cr 
PG1115+080      &4 &16.0   &\cite{weymann80}\cr 
SDSS J1206+4332 &2 &18.5   &\cite{oguri05}\cr 
SDSS J1226$-$0006 &2 &18.3 &\cite{inada03b}\cr 
SDSS J1335+0118 &2 &17.6   &\cite{oguri04b}\cr 
SDSS J1353+1138 &2 &16.5   &\cite{inada06}\cr 
\br
\end{tabular}
\end{indented}
\end{table}

The SQLS has already discovered about 20 new lensed quasars as well as
several previously known lensed quasars. Although the statistical sample
of lensed quasars is still to be finalized, we use these lenses to make
a tentative comparison with the theoretical expectation. To make a fair
comparison with theory, we select a subsample of lenses by choosing
lenses with redshifts $0.6<z<2.2$, magnitudes $i<19.1$, $i$-band flux
ratios (for doubles) $f_i>10^{-0.5}$, image separations
$1''<\theta<3''$, and lens galaxies fainter than the
quasar components $i_{\rm gal}-i_{\rm qso}>0$. Currently we have 13
lensed quasars that meet these conditions, which are summarized in Table
\ref{tab:sdss}. Among these 13 lenses only two are quadruple lenses,
thus the observed quad fraction for the flux limit $i_{\rm lim}=19.1$ is
$p_Q=2/13\simeq0.154$. 

\section{Result}

\begin{figure}
\begin{center}
\includegraphics[width=0.9\textwidth]{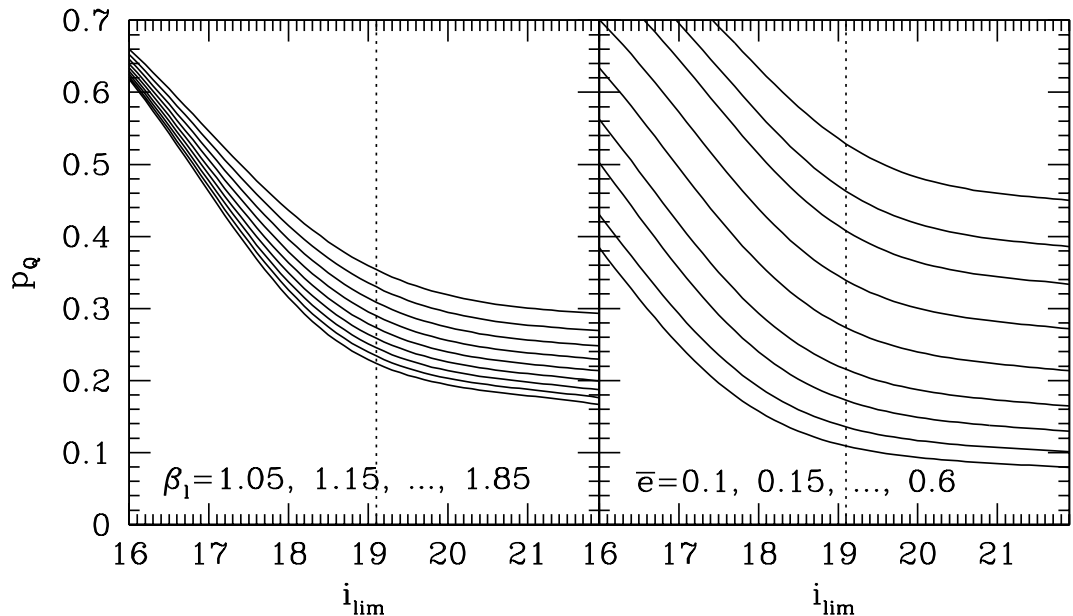}
\end{center}
\caption{The fraction of quadruple lenses $p_Q$ as a function of
 $i$-band limiting magnitude $i_{\rm lim}$. Here we consider lensed
 quasars with redshifts $0.6<z<2.2$, flux ratios $f_i>10^{-0.5}$,
 image separations $1''<\theta<3''$, and lens galaxies fainter than
 the quasar components $i_{\rm gal}-i_{\rm qso}>0$. Dotted 
 line indicate the limiting magnitude of SDSS quasars, $i_{\rm}=19.1$. 
 Left: From lower to upper solid lines, the faint end luminosity
 function of quasars $\beta_{\rm l}$ is changed from $1.05$ to $1.85$. The
 mean ellipticity $\bar{e}$ is fixed to 0.3.
 Right: From lower to upper solid lines, the mean ellipticity $\bar{e}$
 is changed from $0.1$ to $0.5$. The slope $\beta_{\rm l}$ is fixed to 1.45.} 
\label{fig:beta}
\end{figure}

Before comparing our calculation with the observed quad fraction, we
see how it depends on parameters. Among others, the most important
parameter is the ellipticity. Another important element that
determines the quad fraction is the shape of the quasar luminosity
function. In particular the faint end slope $\beta_{\rm l}$ still contains
large errors because current large-scale surveys are not deep enough to
fully explore the faint end luminosity function. For instance,
\cite{boyle00} and \cite{croom04} adopted the 2dF quasar sample to
derive the faint end slopes of $\beta_{\rm l}=1.58$ and $1.09$, respectively.
A survey of faint quasars conducted by \cite{jiang06} suggests that the 
faint end slope could be $\beta_{\rm l}=1.25$, shallower than our
fiducial value. Other uncertainties, such as cosmological
parameters, the velocity function of galaxies, and the number of
source quasars, affect the number of double and quad lenses roughly
similarly, thus they hardly change the fraction of quad lenses. We
find that the effect of changing the prolate/oblate fraction is not
large, affecting the quad fraction only by a few percent. Therefore,
in figure \ref{fig:beta} we plot the quad fraction as a function of
the limiting magnitude $i_{\rm lim}$ changing these two important
parameters, the mean ellipticity $\bar{e}$ and the faint end slope
$\beta_{\rm l}$. First, the quad fraction decreases as the limiting
magnitude increases. Larger magnifications of quads than doubles
indicate that the quad fraction is a strong function of magnification
bias such that larger magnification bias results in larger quad
fraction, which explain the decrease of the quad fraction with
increasing $i_{\rm lim}$. As expected, the quad fraction is quite
sensitive to the ellipticity and the faint end slope of the quasar
luminosity function. 

\begin{figure}
\begin{center}
\includegraphics[width=0.45\textwidth]{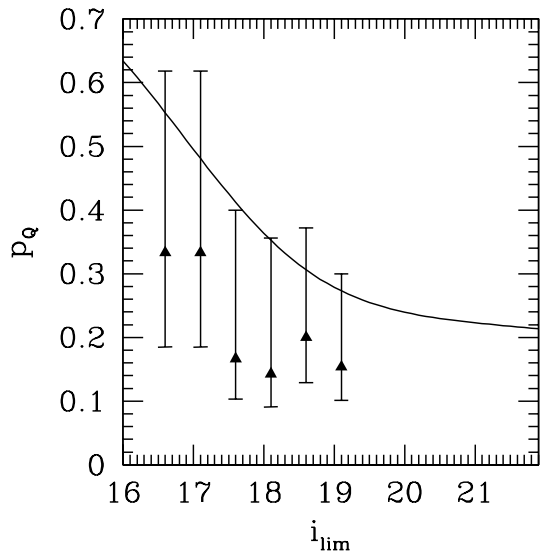}
\end{center}
\caption{The quad fraction in our fiducial model ($\bar{e}=0.3$,
  $\beta_{\rm l}=1.45$; shown by a solid line) is compared with
  observed fractions in the SQLS (filled triangles with errorbars). The
  errors indicate 68\% error estimated assuming the Poisson distribution
  for the numbers of double and quad lenses. See table \ref{tab:sdss}
  for the lens sample we use. Note that the data points are not
  independent but rather correlated in the sense that lenses used to
  plot at each $i_{\rm lim}$ are included in computing data points
  at larger $i_{\rm lim}$ as well.}      
\label{fig:obs}
\end{figure}

Next we compare the quad fraction in our fiducial model with the
observed fraction in the SQLS. Figure \ref{fig:obs} shows both the
theoretical and observed quad fractions as a function of the limiting
magnitude. We find that the observed quad fraction is indeed lower
than the theoretical prediction. For instance, at $i_{\rm lim}=19.1$ 
the quad fraction in our model is $p_Q=0.273$ that is larger than the
observation, $p_Q=0.154$. However, by taking the large errorbar
of the observed fraction due to the small number of lenses, we
conclude that the observed quad fraction is consistent with the
theoretical expectation. 

\begin{figure}
\begin{center}
\includegraphics[width=0.45\textwidth]{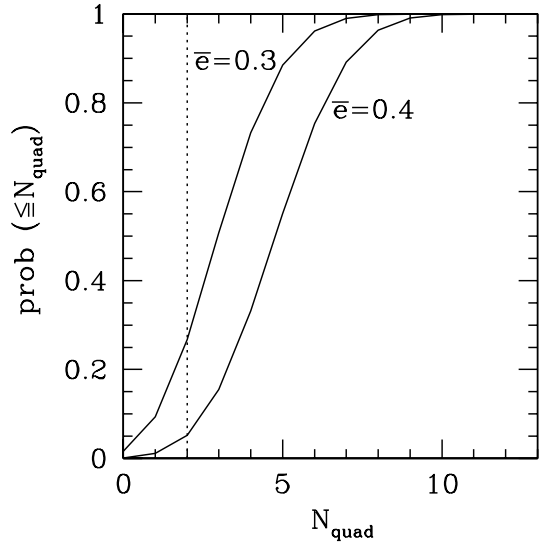}
\end{center}
\caption{Probability of our model producing quad lenses equal or fewer
  than $N_{\rm quad}$ in a sample of 13 lenses, computed from our
  model prediction of the quad fraction for $i_{\rm lim}=19.1$,
  $p_Q=0.273$.  In addition to our fiducial model we also plot the
  probability for $\bar{e}=0.4$ ($p_Q=0.408$) that better reproduces
  the high quad fraction in the CLASS. The observed number of quads in
  the SQLS, $N_{\rm quad}=2$, is indicated by a vertical dotted line.
} 
\label{fig:poi}
\end{figure}

In figure \ref{fig:poi} we plot the probability that our
theoretical model produces $\leq N_{\rm quad}$ quad lenses in a sample
of 13 lenses. Note that in observation there are $N_{\rm quad}=2$ quad
lenses (see table \ref{tab:sdss}). In our fiducial model the
probability is $\simeq 0.27$, which is low but acceptable. On
the other hand, if we increase the mean ellipticity to $\bar{e}=0.4$,
which is roughly the best-fit value for the observed quad fraction in
the CLASS (see \cite{chae03}), the probability reduces to $\simeq
0.05$. Therefore with such large-ellipticity model it is difficult to 
account for the low quad fraction observed in the SQLS.
 
\begin{figure}
\begin{center}
\includegraphics[width=0.9\textwidth]{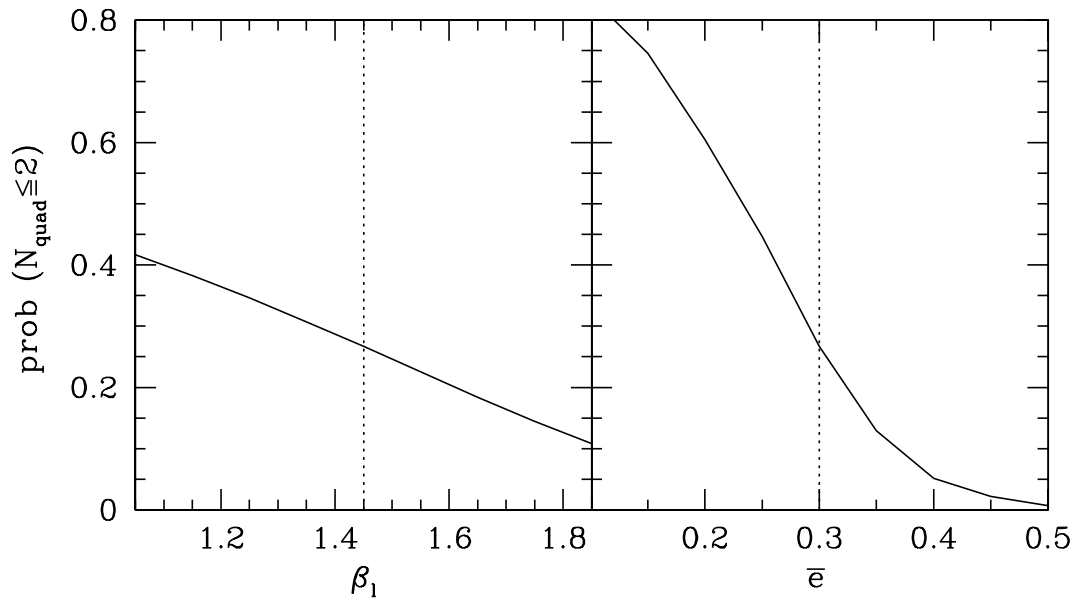}
\end{center}
\caption{Probability of our model producing quad lenses equal or fewer
than the observed case, $N_{\rm quad}=2$, is plotted as a function of
$\beta_{\rm l}$ ({\it left}) or $\bar{e}$ ({\it right}). The fiducial
values are shown by vertical dotted lines.} 
\label{fig:like}
\end{figure}

Finally we check the dependence of the likelihood for $N_{\rm quad}\leq
2$ in a sample of 13 lenses on the faint end slope $\beta_{\rm l}$ and
the mean ellipticity $\bar{e}$ in figure \ref{fig:like}. As expected
from figure \ref{fig:beta}, the probability depends sensitively on
these parameters. For instance, by decreasing the faint end slope to
$\beta_{\rm l}=1.25$, which is preferred by a spectroscopic survey of
faint quasars \cite{jiang06}, the probability is increased to $\simeq
0.35$. Changing the mean ellipticity to $0.2$ enhances the probability
to $\simeq 0.61$, making the observed low quad fraction quite
reasonable.

\section{Summary and Discussion}

In this paper, we have studies the fraction of four-image lenses among
optical gravitational lenses. We have paid a particular emphasis to
whether the low quad fraction observed in the SQLS is consistent with
the standard theoretical prediction. In order to make a fair
comparison, we have taken account of the selection function and source
population in predicting the quad fraction. We find that the observed
quad fraction in the SQLS, $p_Q=2/13\simeq0.154$, is indeed lower than
the prediction of our fiducial model, $p_Q=0.273$, but is consistent
given the large Poisson error of the observed quad fraction. 

We can lower the expected quad fraction by either making the faint end
slope of the quasar luminosity function shallower or decreasing the
mean ellipticity of lens galaxies. However, lowering the ellipticity
decreases {\it both} optical and radio quad fractions, therefore such
models have difficulty in explaining the high quad fraction among CLASS
lenses. For instance, from the CLASS lens sample \cite{chae03} derived
68\% lower limit of the mean ellipticity to 0.28 which is marginally
consistent with our fiducial model, $\bar{e}=0.3$. Therefore, one way
to explain both the high quad fraction among radio lenses and low
quad fraction among optical lenses is to consider a shallow
faint end slope of the quasar optical luminosity function while
keeping the mean ellipticity relatively high.

A caveat is that the SQLS is still ongoing and the lens sample is not
yet finalized. We should use a final, larger lens sample of the SQLS
to draw a more robust conclusion from the quad fraction. The final
statistical sample is expected to contain roughly twice the number of
lenses we used in this paper, thus the statistical error should be 
reduced significantly.

\ack
I thank Naohisa Inada and Kyu-Hyun Chae for discussions, and an
anonymous referee for many suggestions. 
This work was supported in part by the Department of Energy contract 
DE-AC02-76SF00515.

\end{document}